\documentclass[aps,prl,twocolumn,superscriptaddress]{revtex4}
\usepackage{graphicx}
\usepackage{latexsym}
\usepackage{amssymb}
\usepackage{amsmath}
\usepackage{amsfonts}
\usepackage{bm}
\usepackage{multirow}
\usepackage{color}

\newcommand{\bra}[1]{\langle #1|}
\newcommand{\ket}[1]{|#1 \rangle}

\newcommand{\vect}[1]{{\bm{#1}}}
\newcommand{\eqnref}[1]{Eq.\,\eqref{#1}}

\newcommand{\beq}{\begin{equation}}
\newcommand{\eeq}{\end{equation}}
\newcommand{\beqn}{\begin{eqnarray}}
\newcommand{\eeqn}{\end{eqnarray}}

\begin{document}

\title{Topological number and Fermion Green's function of
Strongly Interacting \\ Topological Superconductors}

\author{Yi-Zhuang You}

\affiliation{Department of physics, University of California,
Santa Barbara, CA 93106, USA}

\author{Zhong Wang}

\affiliation{Institute for Advanced Study, Tsinghua University,
Beijing, China, 100084}

\affiliation{Collaborative Innovation Center of Quantum Matter,
Beijing 100871, China}

\author{Jeremy Oon}

\affiliation{Department of physics, University of California,
Santa Barbara, CA 93106, USA}

\author{Cenke Xu}

\affiliation{Department of physics, University of California,
Santa Barbara, CA 93106, USA}

\begin{abstract}

It has been understood that short range interactions can reduce
the classification of topological superconductors in all
dimensions. In this paper we demonstrate by explicit calculations
that when the topological phase transition between two distinct
phases in the noninteracting limit is gapped out by interaction,
the bulk fermion Green's function $G(i\omega)$ at the
``transition" approaches zero as $G(i\omega) \sim \omega$ at
certain momentum $\vec{k}$ in the Brillouin zone.

\end{abstract}

\pacs{}

\maketitle

\emph{Introduction} ---


Unlike bosonic systems, fermionic systems can have nontrivial
topological insulator (TI) and topological superconductor (TSC)
phases without interaction. Noninteracting fermionic TIs and TSCs
in all dimensions with various symmetries have been fully
classified in
Ref.~\onlinecite{ludwigclass1,ludwigclass2,kitaevclass}. However,
it remains an open and challenging question that what role does
interaction play in the classification of TIs and TSCs. According
to a pioneering work Ref.~\onlinecite{fidkowski1,fidkowski2}, a
$1d$ TSC with time-reversal symmetry, which in the noninteracting
limit has a $\mathbb{Z}$
classification~\cite{ludwigclass1,ludwigclass2,kitaevclass}, has
only a $\mathbb Z_8$ classification in the presence of local
interactions. This implies that the boundary state of 8 copies of
this $1d$ TSC (which is 8 flavors of $0d$ Majorana fermion zero
modes) can be gapped out by interaction without ground state
degeneracy. The work in Ref.~\onlinecite{fidkowski1,fidkowski2}
was soon generalized to $2d$ TSCs with a $1d$
boundary~\cite{qiz8,yaoz8,zhangz8,levinguz8}, and $3d$ TSCs with
$2d$ boundary~\cite{chenhe3B,senthilhe3}. All these TSCs have
$\mathbb{Z}$ classification without interaction, namely in all
these cases, for arbitrary flavors of the system, the boundary
remains gapless without interaction. But with 8 copies of such
TSCs in $2d$, and $16$ flavors of He$^3$B in $3d$, a strong enough
local interaction that preserves all the symmetries can render the
boundary states trivial, namely the boundary states can be gapped
out by interaction without degeneracy.

If interaction reduces the classification of a TSC, it not only
implies that the boundary becomes trivial under interaction, it
also implies that the ``TSC" and the trivial state can be
adiabatically connected to each other without any phase
transition, namely the bulk quantum critical point in the
noninteracting limit is gapped out by interaction. A free fermion
TI or TSC with $\mathbb{Z}$ classification is usually
characterized by a quantized topological number, which can be
constructed by fermion Green's
function~\cite{tknn,tknn2,wang1,wang2,wang3,wang4}. For example,
the so-called TKNN number~\cite{tknn,tknn2} of integer quantum
Hall state can be represented as~\cite{tknn3,volovikbook} \beqn N
= \frac{1}{24\pi^2} \int \mathrm{d}^3 k \epsilon^{\mu\nu\rho}
\mathrm{Tr}[G
\partial_\mu G^{-1} G
\partial_\nu G^{-1} G
\partial_\rho G^{-1} ] , \label{tknn}
\eeqn where $G(k)$ is the fermion Green's function in the
frequency and momentum space $k=(i\omega,\vect{k})$, and $i\omega$
is the Matsubara frequency. The formula Eq.~\ref{tknn} can be
formally applied to interacting systems too, as long as we use the
full interacting fermion Green's
function~\cite{tknn,tknn2,wang1,wang2,wang3,wang4}. Recently it
was also proved that in the presence of interaction,
Eq.~\ref{tknn} is fully determined by zero-frequency Green's
function, and the topological invariant takes a simpler
form~\cite{wang3} $ N= \frac{1}{2\pi}\int d^2 k\mathcal{F}_{xy}$,
where $\mathcal{F}_{xy}$ is the Berry curvature calculated from
the eigenvectors of $-G^{-1}(\omega=0, \vect{k})$, as if it is a
``noninteracting'' Hamiltonian. This TKNN topological number must
be quantized, so it can only change through a sharp ``transition"
in the phase diagram. In the noninteracting limit, this sharp
transition is a physical phase transition which corresponds to
closing the bulk gap of the fermion, $i.e.$ the fermion Green's
function develops a gapless pole at the transition $G(i\omega)
\sim 1/\omega$. With strong interaction, the results in
Ref.~\onlinecite{fidkowski1,fidkowski2,qiz8,yaoz8,zhangz8,levinguz8,chenhe3B,senthilhe3}
imply that this quantum critical point can be gapped out by
interactions, while this ``transition" of topological number must
still occur in the phase diagram.
Ref.\,\onlinecite{gurarie,manmana} proposed that the quantized
topological number such as Eq.~\ref{tknn} not necessarily
corresponds to gapless edge states, it can also correspond to
zeros of the fermion Green's function at the edge. In our paper we
will study the fermion Green's function in the bulk, and we will
demonstrate with explicit calculations that the fermion Green's
function develops zero at the ``transition" of topological
numbers, even though there is no transition in bulk spectrum. And
the fermion Green's function vanishes analytically as $G(i\omega)
\sim \omega$.

A simple observation can support this conclusion above: in
\eqnref{tknn}, $G$ and $G^{-1}$ are almost equivalent. 
Thus the topological number should either change through a pole of
$G$ (the noninteracting case), or change through a zero of $G$,
which corresponds to a pole of $G^{-1}$. This implies that when
the ``transition" of topological number is gapped out by
interactions, the Green's function should approach zero as
$G(i\omega) \sim \omega^\alpha$ with positive $\alpha$. Since at
the ``transition" the fermion spectrum is fully gapped, the
fermion correlation function in real space-time must be short
ranged, thus $\alpha$ must be an integer, and the simplest
scenario is $\alpha = 1$.

\emph{$1d$ Example} ---


Before providing a general argument of the existence of zeros in
the fermion Green's function for arbitrary dimension, we will
first demonstrate this behavior explicitly in a $1d$ example using
8 copies of Kitaev's chain\cite{kitaevchain}, also known as the
$1d$ Fidkowski-Kitaev model\cite{fidkowski1}. Consider a $1d$
lattice, on each site $i$, we introduce 8 Majorana fermions
denoted by $\chi_{i\alpha}$ with $\alpha=1,\cdots,8$ labeling the
fermion flavors (species), which satisfy the Majorana fermion
anti-commutation relations
$\{\chi_{i\alpha},\chi_{i'\alpha'}\}=2\delta_{ii'}\delta_{\alpha\alpha'}$.
The model Hamiltonian reads
\begin{equation}\label{eq: FK model}
\begin{split}
H&=H_u + H_w,\\
H_u&=\frac{1}{2}\sum_{i,j,\alpha} iu_{ij}\chi_{i\alpha}\chi_{j\alpha},\\
H_w&=-w\sum_{i,\{\alpha_k\}}X_{\alpha_1\alpha_2\alpha_3\alpha_4}\chi_{i\alpha_1}\chi_{i\alpha_2}\chi_{i\alpha_3}\chi_{i\alpha_4}.
\end{split}
\end{equation}
$H_u$ is the inter-site coupling term with $u_{ij}\in\mathbb{R}$
and $u_{ij}=-u_{ji}$. $H_w$ is the on-site Fidkowski-Kitaev
interaction, in which the coefficient
$X_{\alpha_1\alpha_2\alpha_3\alpha_4}\equiv\bra{e_1}\chi_{\alpha_1}\chi_{\alpha_2}\chi_{\alpha_3}\chi_{\alpha_4}\ket{e_1}$
is given as the expectation value of the four-fermion operator on
a chosen on-site ground state $\ket{e_1}$ (to be specified later).
Here we have omitted the site index $i$, and the summation of
flavor $\sum_\alpha$ in $H_w$ is to sum over all the possible
quartets $\alpha_1<\alpha_2<\alpha_3<\alpha_4$ for
$\alpha_k=1,\cdots,8$. To specify the state $\ket{e_1}$, we may
choose a representation of the Majorana operators as
$\chi_{2n-1}=\sigma_n^1\prod_{m<n}\sigma_m^3$ and
$\chi_{2n}=\sigma_n^2\prod_{m<n}\sigma_m^3$ for $n=1,\cdots,4$,
with $\sigma_n^\mu$ being the $\mu$-th Pauli matrix acting on the
$n$-th tensor factor. Then the chosen ground
state\cite{fidkowski1} can be written as
$\ket{e_1}=(\ket{0000}-\ket{1111})/\sqrt{2}$ (where $\ket{0}$ and
$\ket{1}$ are the eigen states of $\sigma^3$). The choice of
$\ket{e_1}$ is not unique, but we will stick to this convention.
It turns out that there are 14 non-zero
$X_{\alpha_1\alpha_2\alpha_3\alpha_4}$'s, corresponding to the 14
four-fermion terms in $H_w$ on each site. The explicit form of
$H_w$ was given in Ref.\,\onlinecite{fidkowski1}, and its physical
meaning has been discussed in Ref.\,\onlinecite{yaoz8,manmana}.

We would like to introduce an explicit set of eigen basis for $H_w$. Let
us focus on a single site, and omit the site index $i$. In the
16-dimensional Hilbert space of the 8 Majorana fermions on each
site, the Hamiltonian $H_w$ can be diagonalized. All its on-site
many-body eigenstates can be constructed from $\ket{e_1}$ by
applying Majorana fermion operators.  We can divide these states
into even and odd sectors according to their fermion parity. Given
the ground state $\ket{e_1}$ is an even parity state, the odd
parity states $\ket{o_\alpha}=\chi_{\alpha}\ket{e_1}$ are obtained
by acting a single fermion operator; and the even parity states
$\ket{e_\alpha}=\chi_{1}\chi_{\alpha}\ket{e_1}$ can be constructed
by acting the two-fermion operator with the first operator fixed
as $\chi_1$. These states $\ket{p_\alpha}$ ($p=e,o$ and
$\alpha=1,\cdots,8$) form a set of orthonormal basis, i.e.
$\langle
p_\alpha|p'_{\alpha'}\rangle=\delta_{pp'}\delta_{\alpha\alpha'}$.
These states can be defined on each site as $\ket{p_\alpha}_i$. In
this set of basis, the  interaction Hamiltonian $H_w$ is
diagonalized,
\begin{equation}\label{eq: Hint diag}
H_w=\sum_i\Big(-14w\ket{e_1}_i{}_i\bra{e_1}+2w\sum_{\alpha=2}^8\ket{e_\alpha}_i{}_i\bra{e_\alpha}\Big).
\end{equation}
Odd parity states are annihilated by $H_w$, i.e. of zero
eigenvalue, so they do not appear in \eqnref{eq: Hint diag}. The
many-body ground state of $H_w$ is simply the $\ket{e_1}$ product
state, denoted as $\ket{0}=\prod_{i}\ket{e_1}_i$, with the ground
state energy $E_0$ scaling with the system size.

Now we turn on the bilinear fermion coupling term $H_u$
perturbatively, assuming $u_{ij}\ll w$. In this section we will
calculate the Green's function up to the first order perturbation
of $u_{ij}/w$. In the next section we will prove that our
qualitatively result is valid after summing over the entire
infinite perturbation series. $H_u$ will take the unperturbed
ground state $\ket{0}$ to a group of two-fermion excited  states
$\chi_{i\alpha}\chi_{j\alpha}\ket{0}$, all of which have
approximately the same energy $E_0+28w$. So according to the
perturbation theory, the perturbed ground state (to the first
order in $u_{ij}$) reads
\begin{equation}
\ket{g}=\Big(1-\frac{H_u}{28w}\Big)\ket{0}.
\end{equation}
The Majorana fermion Green's function in the real space can be
calculated from\cite{wang3,wang4}
\begin{equation}
\begin{split}
G(i\omega)_{i\alpha,i'\alpha'}=\sum_{m}\Big(&
\frac{\bra{g}\chi_{i\alpha}\ket{m}\bra{m}\chi_{i'\alpha'}\ket{g}}{i\omega-(E_m-E_0)}\\
&+\frac{\bra{m}\chi_{i\alpha}\ket{g}\bra{g}\chi_{i'\alpha'}\ket{m}}{i\omega+(E_m-E_0)}\Big).\\
\end{split}
\end{equation}
Since $H_u\ll H_w$, we have $(E_m-E_0)\simeq 14w$, therefore
\begin{equation}\label{eq: G spectral representation}
\begin{split}
G(i\omega)_{i\alpha,i'\alpha'}=&\frac{\bra{g}\chi_{i\alpha}\chi_{i'\alpha'}\ket{g}}{i\omega-14w}+\frac{\bra{g}\chi_{i'\alpha'}\chi_{i\alpha}\ket{g}}{i\omega+14w}\\
=&\frac{i\omega}{(i\omega)^2-(14w)^2}\bra{g}\{\chi_{i\alpha},\chi_{i'\alpha'}\}\ket{g}\\
&+\frac{14w}{(i\omega)^2-(14w)^2}\bra{g}[\chi_{i\alpha},\chi_{i'\alpha'}]\ket{g}.
\end{split}
\end{equation}
In the first term,
$\bra{g}\{\chi_{i\alpha},\chi_{i'\alpha'}\}\ket{g}=2\delta_{ii'}\delta_{\alpha\alpha'}$
according to the Majorana fermion anti-commutation relations. The
second term is not vanishing only for $i\neq i'$, and under such
condition, $
\bra{g}[\chi_{i\alpha},\chi_{i'\alpha'}]\ket{g}=2\bra{g}\chi_{i\alpha}\chi_{i'\alpha'}\ket{g}=4\bra{0}\chi_{i\alpha}\chi_{i'\alpha'}\frac{H_u}{-28w}\ket{0}=2iu_{ii'}\delta_{\alpha\alpha'}/(14w)$.
Hence
\begin{equation}\label{eq: G real space}
G(i\omega)_{i\alpha,i'\alpha'}=\frac{2\delta_{\alpha\alpha'}}{(i\omega)^2-(14w)^2}(i\omega\delta_{ii'}+iu_{ii'}).
\end{equation}
Fourier transform to the momentum space, assuming the coupling
$u_{ij}$ is alternating between $u$ and $v$ along the $1d$ chain,
s.t. $u_\vect{k}=u-ve^{i\vect{k}}$, then on the sublattice basis,
\begin{equation}
G(i\omega,\vect{k})_{\alpha,\alpha'}=
\frac{2\delta_{\alpha\alpha'}}{(i\omega)^2-(14w)^2} \left[
\begin{matrix}
i\omega & -iu_\vect{k}^*\\
iu_\vect{k} & i\omega
\end{matrix}
\right]. \label{green}
\end{equation}
At the topological ``transition'' point $u=v$, $u_\vect{k}\sim -iv
\vect{k}$ vanishes as $\vect{k}\to 0$, so the Green's function
approaches zero analytically at the zero frequency and momentum
$i\omega,\vect{k}\to0$ as expected.
In contrast, in the free fermion case, such a topological
transition happens through a gapless critical point characterized
by the pole in the Green's function instead.

\emph{Zero in Green's Function in general dimension} ---


The above perturbative calculation is not limited to the $1d$
model, but can be immediately generalized to higher spatial
dimensions. It is found that for a series of $d$-dimensional
models with arbitrary $d$, the topological number defined with
full fermion Green's function can change via the zero in the
Green's function at some momentum in the Brillouin zone. This
statement actually holds to all orders of perturbation.



We will consider lattice models for a $d$-dimensional TSC ($d \geq
2$) like the follows:
\begin{equation}
\begin{split}
H =&\sum_{\alpha = 1}^8 \chi^\intercal_{\alpha,-\vect{k}}
\Big(\sum_{\mu = 1}^d \sin k_\mu \Gamma^\mu \Big) \chi_{\alpha,\vect{k}} \\ &+m \chi^\intercal_{\alpha, -\vect{k}} \Big( \sum_{\mu = 1}^d \cos
k_\mu - d + 1 \Big) \Gamma^{d+1} \chi_{\alpha, \vect{k}}; \label{dtsc}
\end{split}
\end{equation}
where $\Gamma^{1 \cdots d}$ are symmetric matrices, while
$\Gamma^{d+1}$ is an antisymmetric matrix. When $d = 2$, $\Gamma^1
= \sigma^x$, $\Gamma^2 = \sigma^z$, $\Gamma^3 = - \sigma^y$; $m =
0$ is the quantum critical point between 8 copies of $p + ip$ TSC
($m
> 0$) and $p - ip$ TSC ($m < 0$), the topological number defined
in \eqnref{tknn} changes by 16 at this transition. When $d = 3$,
$\Gamma^{1,2,3}$ are $4 \times 4$ symmetric matrices, $\Gamma^4$
and $\Gamma^5 = \prod_{a = 1}^4 \Gamma^a$ are antisymmetric
matrices. $m = 0$ is the quantum critical point between 8 copies
of He$^3$B TSC with topological number $-1$ and $+1$. The
time-reversal symmetry acts on the Majorana fermions as $Z_2^T:
\chi_{\alpha, \vect{k}} \rightarrow i \Gamma^5 K \chi_{\alpha,
-\vect{k}}$, where $K$ stands for complex conjugation. The
topological number for He$^3$B phase can be represented with
Fermion Green's function: \cite{volovik2011,wang1,wang3} \beqn
\label{3dN}
N=\frac{1}{48\pi^2}\int\mathrm{d}^3k\epsilon^{abc}\mathrm{Tr}[\Gamma^5
G\partial_a G^{-1} G\partial_b G^{-1} G\partial_c G^{-1}], \eeqn
where $G = G(0, \vect{k})$ is the zero frequency Green's function.
The momentum integral is carried out in the Brillouin zone.

Besides the Majorana fermion hopping terms, we will also turn on
the on-site Fidkowski-Kitaev interaction $H_w$ to gap out the
fermions,
\begin{equation}
H_w=-w \sum_{j}
\sum_{\sigma,\{\alpha_k\}}X_{\{\alpha_k\}}\chi_{\sigma\alpha_1,
j}\chi_{\sigma\alpha_2, j}\chi_{\sigma\alpha_3,
j}\chi_{\sigma\alpha_4, j}, \label{w}
\end{equation}
where $\sigma$ labels the orbital degrees of freedom, $j$ labels
the lattice sites. The on-site interaction in \eqnref{w} has a
nondegenerate ground state on every site, thus with strong
interaction \eqnref{w}, the quantum critical point $m = 0$ in
\eqnref{dtsc} is indeed gapped, and the system is in a trivial
direct product state.

Now we will demonstrate that in the two dimensional phase diagram
of $w$ and $m$, $m = 0$ is always the transition line of
topological number defined with the Green's function
(Fig.~\ref{fig: phase diagrams}). With weak interaction $w$,
because the topological number defined in Eq.~\ref{tknn} and
Eq.~\ref{3dN} are always quantized, a weak interaction will not
change the topological numbers. With strong interaction, the
fermion Green's function can be computed by perturbation theory as
the previous section. With first order perturbation, the Green's
function in any dimension would take the same form as
Eq.~\ref{green}. It is straightforward to verify that the strongly
interacting Green's function from first order perturbation theory
always leads to the same $2d$ TKNN number (Eq.~\ref{tknn}) as 8
copies of noninteracting $p \pm ip$ superconductors with the same
$m$. Also, for strongly interacting He$^3$B, according to
Eq.~\ref{green}, the zero frequency Greens function reads $G(0,
\vect{k}) \sim 1/ G(0, \vect{k})_{0}$, where $G(i\omega,
\vect{k})_{0}$ is the free fermion Green's function without
interaction at all. In the definition of topological number
Eq.~\ref{3dN}, $G(0,\vect{k})$ and $G(0,\vect{k})^{-1}$ are
totally interchangeable since $G(0,\vect{k})$ anticommutes with
$\Gamma_5$, thus the topological numbers in the noninteracting
limit and strong coupling limit are identical. This means that $m
= 0$ is always the ``transition" line of topological number, even
though the qualitative spectrum of the system might not change
across this line. The argument above was based on the explicit
form of the fermion Green's function from the first order
perturbation theory. But since the topological numbers are always
quantized, we expect the higher order perturbations will not
change the topological numbers, as long as $w$ is strong enough.

Now we analyze the symmetry of the two-fermion Green's function
$G(k)_{\alpha,\alpha'}=-\langle
\chi_{k\alpha}\chi_{-k\alpha'}^\intercal\rangle$ where
$\chi_{k\alpha}$ denotes the $\alpha$-flavor Majorana fermion of
the frequency-momentum $k=(i\omega,\vect{k})$. In the flavor space
(indexed by $\alpha,\alpha'$), $H_u$ has the full SO(8) symmetry
and $H_w$ has a smaller SO(7) symmetry\cite{fidkowski1}, so the
total Hamiltonian $H=H_u+H_w$ is SO(7) symmetric, and
$\chi_\alpha$ carries a 8-dimensional spinor representation of the
SO(7) symmetry. Since with strong interaction \eqnref{w} the
system is gapped and nondegenerate, the ground state must be SO(7)
invariant. Thus the Green's function must be diagonal in the
flavor space: $G(k)_{\alpha,\alpha'}=g(k)\delta_{\alpha\alpha'}$,
where $g(k)$ is the Green's function in the orbital space (see
Appendix A for a mathematical proof). Physically this can be
understood since $H_w$ only breaks the SO(8) symmetry by driving a
four-fermion ordering without generating any fermion-bilinear
order, so the broken symmetry will not be revealed at the
two-fermion level. Then we turn to the orbital space Green's
function $g(k)$. Because the Fidkowski-Kitaev interaction $H_w$
has the full orbital space symmetry, in the strong interaction
limit, the orbital degrees of freedom must be degenerate at each
energy level. So the Green's function can be written as a
polynomial of $H_u/(i\omega - H_w)$ where $H_w$ will just behave
like a constant in the orbital space. Given that at the
``transition'' point $H_u$ is constructed using symmetric matrices
$\Gamma^\mu$ only, so the Green's function $g(k)$ must also be a
symmetric matrix (in the momentum space), i.e.
$g(k)=g(k)^\intercal$. However, by definition $g(k)=-\langle
\chi_{k}\chi_{-k}^\intercal\rangle$, and for the off-diagonal part
of $g(k)$, we have $g(-k)=-\langle
\chi_{-k}\chi_{k}^\intercal\rangle=\langle
\chi_{k}\chi_{-k}^\intercal\rangle^\intercal=-g(k)^\intercal=-g(k)$,
meaning that the off-diagonal part of $g(k)$ must be odd in $k$.
Since in the large $w$ limit the whole system is gapped, the real
space fermion Green's function must decay exponentially, thus the
off-diagonal components of $g(k)$ must be analytic when $ k
\rightarrow 0$. Thus we conclude that the off-diagonal components
of $g(k)$ must scale as $g(k) \sim k^{\alpha}$, where $\alpha$
must be a positive odd integer.

So now we are left with the diagonal part of $g(k)$. Using the
spectral representation\cite{wang3} of the Green's function, we
can show that the diagonal part of $g(k)$ at zero-momentum must
take the following form
\begin{equation}
g(i\omega,\vect{k}=0)_{\sigma,\sigma}=\sum_{m}\frac{2i\omega\;|\bra{m}\chi_{0\sigma}\ket{g}|^2}{(i\omega)^2-(E_m-E_g)^2},
\end{equation}
where $\ket{g}$ is the ground state, and $\ket{m}$ denotes a
generic excited state. We can see
$g(i\omega,\vect{k}=0)_{\sigma,\sigma}$ must be odd in $i\omega$
at $\vect{k}=0$. As long as we are in the fully gapped phase
(under strong enough interaction), the excitation energy
$(E_m-E_g)$ will never vanish, so there will be no singularity at
zero frequency, then it follows that diagonal term of the Green's
function must also vanish with $k\to0$. So in conclusion, the
fermion Green's function at the topological ``transition'' in the
strong interaction phase must develop a zero at zero frequency
around the free-fermion Dirac points in the Brillouin zone.

\emph{Summary and Discussion}
---

\begin{figure}
\begin{center}
\includegraphics[width=240pt]{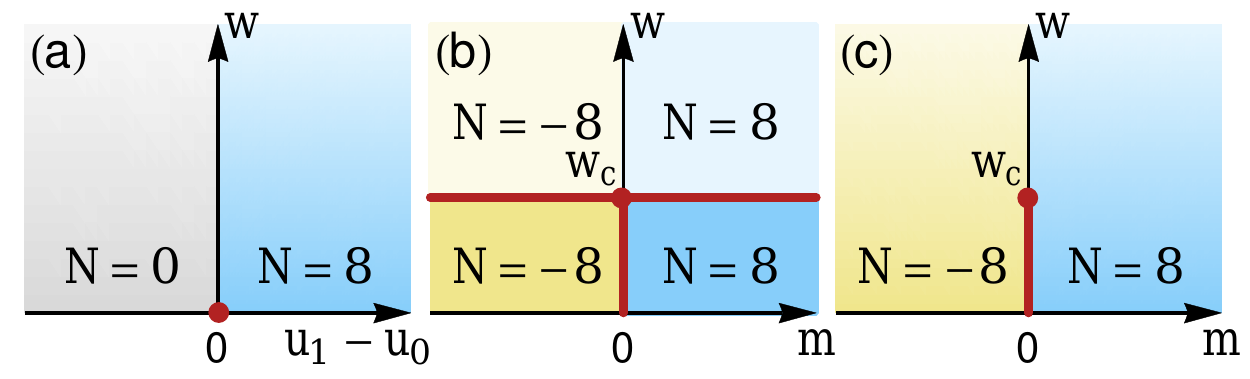}
\caption{Schematic phase diagrams in (a) $1d$, (b) $2d$ and (c)
$3d$. Red line/point marks out the physical phase transition
line/point, where the fermion becomes gapless. The background
color indicates the topological number.} \label{fig: phase
diagrams}
\end{center}
\end{figure}

Let us summarize our results in Fig.~\ref{fig: phase diagrams}. In
all dimensions we have considered in this paper, the $m = 0$ line
in the phase diagrams are always the transition lines for
topological numbers. And in all dimensions a short range
interaction can gap out this transition line, but the fermion
Green's function on this line develops zero at zero frequency.


In $1d$, the system we consider is a 8 copies of Kitaev's chain
with interactions. In $1d$, the quantum critical point at the
noninteracting limit will be immediately gapped out by
infinitesimal interactions, but in $2d$ and $3d$ a short range
interaction is irrelevant for gapless Dirac and Majorana fermions,
thus the interaction can only gap out the critical line $m = 0$
beyond certain value $w_c$. The $2d$ example of our paper is a
transition between eight $p + ip$ to eight $p - ip$ TSCs, and
since these two phases have opposite chiral edge states (opposite
thermal Hall effects), thus they cannot be adiabatically connected
to each other in the phase diagram. Thus although the $m = 0$ line
is gapped out by strong enough interactions, there must be another
phase transition line in the phase diagram (the horizontal line in
Fig.~\ref{fig: phase diagrams}$b$) that separates the strongly
interacting phase (which is a trivial direct product state) from
the $p+ip$ and $p-ip$ TSCs. In $1d$ and $3d$ cases discussed in
our paper, the two sides across the line $m = 0$ can be
adiabatically connected to each other without any physical phase
transition, $i.e.$ the entire phase diagram has only one phase.

Our result implies that when the topological number such as
\eqnref{tknn} is nonzero, the system can still remain trivial.
Ref.~\onlinecite{scorrelator} proposed a different method to
diagnose TIs and TSCs in $1d$ and $2d$: if $|\Psi\rangle$ is a
nontrivial TI or TSC, and $|\Omega \rangle$ is a trivial state,
then the following quantity, so called strange correlator, must
either saturate to a constant or decay as power-law in the limit
$|r - r^\prime| \rightarrow +\infty$: \beqn C(r, r^\prime) =
\frac{\langle \Omega | c(r) \ c^\dagger(r^\prime) | \Psi
\rangle}{\langle \Omega | \Psi\rangle}, \eeqn even though
$|\Omega\rangle$ and $|\Psi\rangle$ both only have short range
correlation between fermion operators. In another work we will
demonstrate that the strange correlator proposed in
Ref.~\onlinecite{scorrelator} can still correctly distinguish
trivial states from nontrivial TSCs under interaction. In
particular, we calculate $C(r, r^\prime)$ for eight copies of $1d$
Kitaev's chain\cite{kitaevchain}. In the TSC phase, $C(r,
r^\prime)$ is a constant in the noninteracting limit, while $C(r,
r^\prime)$ immediately becomes short range with infinitesimal
short range interactions~\cite{youfuture}.

Cenke Xu and Yizhuang You are supported by the David and Lucile
Packard Foundation and NSF Grant No. DMR-1151208, Zhong Wang is
supported by NSFC under Grant No. 11304175, Jeremy Oon is funded
by the NSS Scholarship from the Agency of Science, Technology and
Research (A*STAR) Singapore. The authors would like to thank
Shou-Cheng Zhang for helpful discussions.


\bibliography{zero}

\onecolumngrid
\appendix

\section{Explicit Verification of the Emergent SO(8) Symmetry}
In the fermion flavor space, the Hamiltonian $H=H_u+H_w$ only has
an SO(7) symmetry, but the resulting two-fermion Green's
function actually has an emergent SO(8) symmetry, which is
larger than the symmetry of the Hamiltonian. The smaller SO(7)
symmetry will be manifested in the four-fermion (and higher order)
Green's function, while at the two-fermion level, the Green's
function still processes the full SO(8) symmetry.

To verify this statement, we note that the 8 flavors of the
Majorana fermions form the 8-dimensional spinor representation of
the SO(7) group. We may write down the 21 generators of the
SO(7) group in this spinor representation explicitly. Let
$T_{ij}$ be the generator that performs the SO(7) rotation in
the $ij$-plane. Its spinor representation is an $8\times8$ matrix,
which can be written as a direct product of three Pauli matrices,
denoted as
$\sigma^{\mu\nu\lambda}\equiv\sigma^\mu\otimes\sigma^\nu\otimes\sigma^\lambda$,
\begin{equation}
\begin{array}{ccccccc}
T_{12}=i \sigma^{321} & T_{13}=-i \sigma^{023} & T_{14}=i \sigma^{201} & T_{15}=-i \sigma^{233} & T_{16}=i \sigma^{121} & T_{17}=-i \sigma^{213} & T_{23}=i \sigma^{302}\\
T_{24}=-i \sigma^{120} & T_{25}=i \sigma^{112} & T_{26}=i \sigma^{200} & T_{27}=-i \sigma^{132} & T_{34}=-i \sigma^{222} & T_{35}=i \sigma^{210} & T_{36}=-i \sigma^{102}\\
T_{37}=-i \sigma^{230} & T_{45}=i \sigma^{032} & T_{46}=-i \sigma^{320} & T_{47}=i \sigma^{012} & T_{56}=i \sigma^{312} & T_{57}=i \sigma^{020} & T_{67}=i \sigma^{332}
\end{array}.
\end{equation}
To respect the SO(7) symmetry, the two-fermion Greens function
$G$ must commute with all the 21 generators listed above. We
search over the space of all $8\times8$ matrices by a Mathematica
program, and found that $\sigma^{000}$ (the $8\times8$ identity
matrix) is the only matrix that commute with all the SO(7)
generators. So we must have $G\propto \sigma^{000}$ (or
$G_{\alpha,\alpha'}\propto \delta_{\alpha\alpha'}$) in the fermion
flavor space, which indeed has the full SO(8) symmetry.

\end{document}